# LitCovid in 2022: an information resource for the COVID-19 literature

Qingyu Chen[1,†], Alexis Allot[1,†], Robert Leaman[1], Chih-Hsuan Wei[1], Elaheh Aghaarabi[1,2,§], John Guerrerio[1,3,§], Lilly Xu[1,4,§], Zhiyong Lu[1,*]

1 National Center for Biotechnology Information, National Library of Medicine, National Institutes of Health, MD, USA

2 Towson University, Towson, MD, USA

3 Gilman School, MD, USA

4 Land O' Lakes High School, FL, USA

† The authors wish it to be known that, in their opinion, the first two authors should be regarded as Joint First Authors.

§ The work was completed during the internship.

* To whom correspondence should be addressed. Tel: +1 301 594 7089; Fax: +1 301 480 2290; Email: zhiyong.lu@nih.gov

## ABSTRACT

LitCovid (https://www.ncbi.nlm.nih.gov/research/coronavirus/) – first launched in February 2020 – is a first-of-its-kind literature hub for tracking up-to-date published research on COVID-19. The number of articles in LitCovid has increased from 55,000 to ~300,000 over the past two and half years, with a consistent growth rate of ~10,000 articles per month. In addition to the rapid literature growth, the COVID-19 pandemic has evolved dramatically. For instance, the Omicron variant has now accounted for over 98% of new infections in the U.S. In response to the continuing evolution of the COVID-19 pandemic, this article describes significant updates to LitCovid over the last two years. First, we introduced the Long Covid collection consisting of the articles on COVID-19 survivors experiencing ongoing multisystemic symptoms, including respiratory issues, cardiovascular disease, cognitive impairment, and profound fatigue. Second, we provided new annotations on the latest COVID-19 strains and vaccines mentioned in the literature. Third, we improved several existing features with more accurate machine learning algorithms for annotating topics and classifying articles relevant to COVID-19. LitCovid has been widely used with millions of accesses by users worldwide on various information needs and continues to play a critical role in collecting, curating, and standardizing the latest knowledge on the COVID-19 literature.

## INTRODUCTION

The COVID-19 pandemic has resulted in over 605 million confirmed cases and 6 million deaths globally as of September 2022 (https://covid19.who.int/). Over the past two years, scientists and healthcare professionals worldwide have made significant progress toward understanding its disease mechanisms, prevention, and treatments. Meanwhile, the number of new scholarly articles on COVID-19 has been increasing steadily – approximately 10,000 per month since May 2020 – and has accounted for ~9% of new articles in the entire PubMed (1). This creates significant information overload to keep up with the latest SARS-CoV-2 and COVID-19 research.

In early 2020, we launched LitCovid (https://www.ncbi.nlm.nih.gov/research/coronavirus/), the first-of-its-kind literature hub for tracking up-to-date published research on COVID-19 (2,3). It is updated daily with a combination of machine learning assisted annotation and manual review. When first published in 2020, LitCovid tracked 55,000 relevant articles (3). As of September 2022, not only has the number of COVID-19 articles grown by more than four times, but the COVID-19 pandemic itself has also evolved substantially. Both our understanding of the virus and the properties of the virus itself are significantly different compared with two years ago. One particular instance is the long-term symptoms experienced by a significant percentage of COVID-19 survivors, a condition named Long COVID by the patients affected (4). Some survivors of acute COVID-19 began reporting symptoms lasting much longer than the amount of time then reported for clinical recovery. Long COVID has caused skepticisms and slow responses, and to date, there is no effective treatment (5). In contrast, there is now substantial evidence that a significant percentage of COVID-19 survivors experiencing ongoing multisystemic symptoms (6-9), including respiratory issues (10), cardiovascular disease (11), cognitive impairment (12), and profound fatigue (13). Another significant difference compared with 2020 is the rapid evolution of COVID-19 variants and the deployment of COVID-19 vaccines. Prominent variants such as Alpha, Delta, and more recently Omicron have changed the virus properties and may affect the performance of diagnostic tools and vaccines. According to the report on 3$^{rd}$ September 2022 by the Centers for Disease Control and Prevention (CDC) in the United States (US), the Omicron B4.5 variant accounts for over 88% of COVID-19 infections in the US (https://covid.cdc.gov/covid-data-tracker/#variant-proportions). Amid the evolution of COVID-19 variants, over 700 clinical trials have been dedicated to COVID-19 vaccine candidate development and validation (https://covid19.trackvaccines.org/). To date, World Health Organization (WHO) has 12 COVID-19 vaccines for emergency use (https://covid19.trackvaccines.org/agency/who/); CDC recommended the first updated COVID-19 booster specifically designed to combat Omicron variants on September 1$^{st}$ 2022 (https://www.cdc.gov/media/releases/2022/s0901-covid-19-booster.html). Therefore, tracking COVID-19 variants and vaccines is critical for the ongoing response to the COVID-19 pandemic.

This article describes our continuous efforts to enhance LitCovid in response to these recent developments. First, we created the Long Covid collection in LitCovid to identify biomedical research articles useful for researchers, clinicians, and patients via a human-in-the-loop machine learning approach. Second, we began annotating the latest COVID-19 strains and vaccines mentioned in the literature based on WHO and other authenticated resources. Third, several existing features have also been improved with more accurate and competition-winning deep learning algorithms for annotating topics and identifying COVID-19 relevant articles, respectively. Finally, a new web interface has been developed to improve the user experience.

## METHODS

The main curation workflow of LitCovid is described in its initial publication in 2020 (3). Every day, new articles from PubMed are classified first regarding their relevance to COVID-19. Then, we annotate relevant COVID-19 topics (e.g., treatment) and extract key entities (e.g., vaccine names) mentioned in the literature. Machine learning models have been developed, evaluated, and incrementally updated to relieve the burden of manual review. Herein we describe significant updates in the past two years.

## Classifying articles on Long Covid

A significant update is the annotation of Long COVID articles. Compared with the initial LitCovid publication in early 2020, we now know that a significant percentage of COVID-19 survivors experience ongoing multisystemic symptoms that often affect daily living, a condition known as Long COVID or post-acute-sequelae of SARS-CoV-2 infection (14,15). Identifying Long COVID articles is challenging because this is still a topic under research; studies use a variety of less common terms to describe the condition and most articles do not refer to the condition by name. We developed an iterative human-in-the-loop machine learning framework to efficiently use manually curated labels and also prioritize the documents that need manual review (16). For each iteration of the loop, the framework takes eight input signals including the output from a Long COVID mention recognizer, predictions from LitSuggest machine learning models (17), entity annotations from PubTator (18), and other resources and provides final predictions for manual review. The related data and documentation are publicly available via https://ftp.ncbi.nlm.nih.gov/pub/lu/LongCovid/.

## Annotating Covid-19 Variants and Vaccines

We manually annotated a benchmark dataset of 500 articles in LitCovid with a total of 2,456 COVID-19 variant (e.g., Beta, B.1.351) and vaccine mentions (e.g., mRNA-1273) and subsequently developed and evaluated automatic named entity recognition (NER) methods. A team of five annotated three entity types: COVID-19 virus strains (e.g., Beta, B.1.351), vaccines (e.g., mRNA-1273), and vaccine funders (e.g., Moderna) (the companies or the affiliations providing the vaccines) and normalized them into the corresponding concepts. For example, we link the vaccine mRNA-1273 to its vaccine funder (i.e., Moderna). All five curators annotated the first 100 samples and achieved an inter-annotator agreement of 80.5% (the ratio of exact matches (text span, entity type, and normalized concepts) out of the total number of entities). The remaining documents were further annotated in batches (two annotators per article) and the conflicts were addressed by a third annotator.

The benchmark set enables the development and validation of our NER methods on COVID-19 variants and vaccines. Our NER method consists of two components: entity recognition (identifying whether a text mention is an entity of interests) and entity normalization (normalizing entities into corresponding concepts). For recognition, we applied pre-trained PubMedBERT (3) and fine-tuned on the manually curated dataset. The dataset is randomly split into 400 articles for training and 100 articles for testing. For normalization, we developed a dictionary-lookup approach using the lexicon collected by multiple resources (e.g., MeSH and CDC). The method will also be incrementally updated with the manual annotations on new articles in LitCovid.

## Improvements in document triage and topic annotation

*Document triage*. Document triage identifies whether a new PubMed article is related to COVID-19; if so, the document will be added to LitCovid and annotated further. We have observed an increasing number of articles that mention such keywords as COVID-19 but are not about COVID-19 research. For instance, a study mentions COVID-19 only as a background description and is actually on tuberculosis ("Tuberculosis is caused by the bacterium Mycobacterium tuberculosis (Mtb) and is ranked as the second killer infectious disease after COVID-19" (PMID35926511)). Such cases not only increase the curation workflow but also bring inconsistencies to downstream users. We categorized these articles mentioning COVID-19 but are not related to COVID-19 into three primary categories: (1) the results or findings are not related to COVID-19 (e.g., a study mentions its analysis was completed before the COVID-19 pandemic (PMID35203081)), (2) studies only introduce COVID-19 as a background without stating or quantifying the impacts of COVID-19 (e.g., the Tuberculosis example (PMID35926511)), and (3) studies mention COVID-19 in other fields rather than the main text (e.g., COVID-19 is mentioned in the funding statement (PMID 36044171). We have been manually identifying hundreds of such cases and included them into our training data for calibrating automated triage method.

*Topic annotation*. Topic annotation assigns a set of possible topics (e.g., Treatment and Diagnosis) to an article in LitCovid. Previously we used binary classification models where each model predicts a single topic. In contrast, we have employed LITMC-BERT, a transformer-based multi-label classification method (19). Compared with the previous topic annotation model deployed in LitCovid, it better captures the correlations between topics and provides all the topic predictions simultaneously via multi-task learning.

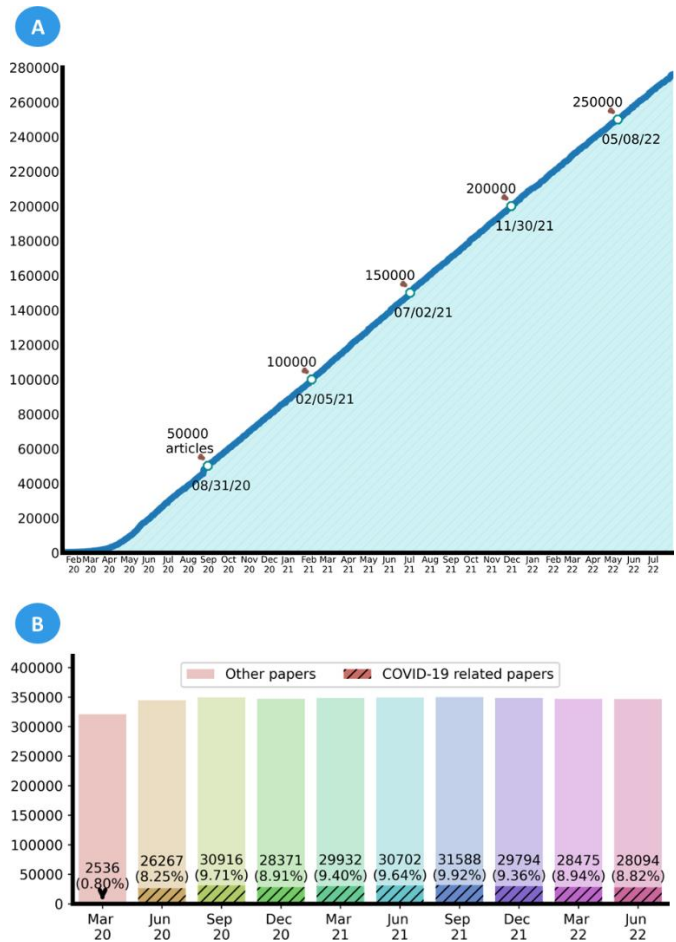

Figure 1. The overall growth of LitCovid by July 2022. (A): the accumulated daily literature growth; (B): the quarterly ratio of LitCovid articles compared with entire PubMed articles

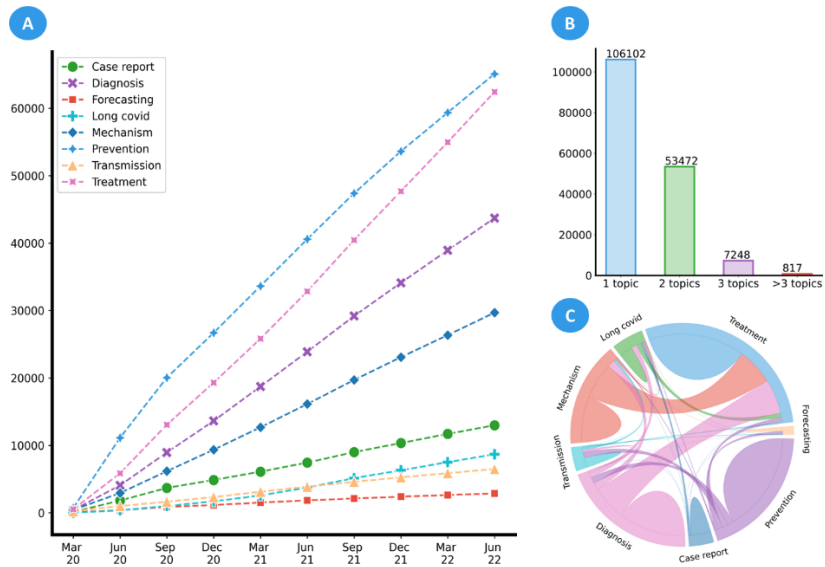

Figure 2. The overall growth of LitCovid topics by July 2022. (A): the accumulated topic growth; (B) the distributions of topics assigned per LitCovid article; (C): topic co-occurrences.

## RESULTS

### Literature growth

Compared with the initial LitCovid publication in October 2020, the number of articles in LitCovid has increased by over five times – from 55,000 to over 280,000. Figure 1 illustrates the overall growth of LitCovid by July 2022. The number of articles has been growing rapidly and consistently with ~10,000 new articles every month for over two years (Figure 1 (A)). Since the second quarter of 2020, the number of LitCovid articles has accounted for ~9% of the new articles in the entire PubMed (Figure 1 (B)). The articles in LitCovid are annotated with up to eight broad topics when applicable, including the new topic Long COVID, which will be detailed later. Figure 2 demonstrates the overall growth of these topics by July 2022. Treatment and Prevention have been the two largest topics, each with over 60,000 articles; in addition, the number of articles with the Long COVID topic has increased more rapidly since March 2021 (Figure 2 (A)). More than half of the articles in LitCovid have been annotated with one or two topics (Figure 2(B)). The new Long COVID topic has been frequently co-occurred with Diagnosis (e.g., representative symptoms of Long COVID (PMID36004306)), Treatment (e.g., therapies for Long COVID treatment (PMID35632649)), and Case Report (e.g., specific Long COVID patient cases (PMID35366017)) (Figure 2(C)).

Table 1. Training and testing datasets and evaluation results of automation tools

| Modules | Train/Test | Precision/Recall/F1-score |
| --- | --- | --- |
| Strains and vaccines recognition | 400/100 | 99.4/98.8/98.6 |
| Strains and vaccines normalization | NA/500 | 99.6/98.4/99.0 |
| Long COVID | 5,733/1,911 | 77.0/87.7/82.0 |
| Topic annotations | 31,199/2,500 | 93.7/91.5/92.1 |
| Document triage (overall) | 39,975/10,025 | 99.5/95.4/97.4 |

## Evaluation performance of automated methods in curation assistance

We evaluated all the automated methods described above before their first use and have improved them continuously. Table 1 summarizes the current performance; all the evaluation datasets and documentation are also publicly available via https://ftp.ncbi.nlm.nih.gov/pub/lu/LitCovid/LitCovid2/litcovid_evaluation/.

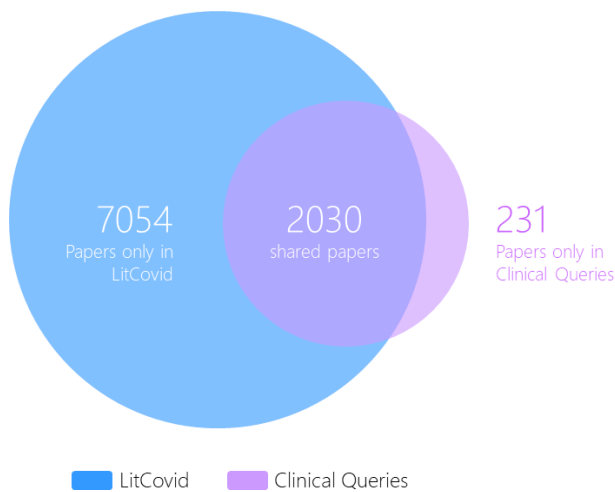

Figure 3. Comparative analysis of the Long COVID topic coverage between LitCovid and Clinical Queries.

*Long COVID Collection.* As demonstrated in Table 1, the method achieved an F1-score of 0.82 for identifying articles pertinent to Long Covid. The annotation of Long COVID articles has been deployed in LitCovid and has been continuously updated. LitCovid has so far accumulated almost 10,000 Long COVID articles (https://www.ncbi.nlm.nih.gov/research/coronavirus/docsum?filters=e_condition.LongCovid ) and nearly 70% of them do not mention Long COVID by name in the title or abstract. We have further collected over 800 synonymous terms for Long COVID and shared to the community. We further compared the Long COVID topic coverage between LitCovid and PubMed Clinical Queries (20) up to July 2022. PubMed Clinical Queries used the eight topic names in LitCovid but is based on keyword matching (20). Figure 3 shows the Long COVID topic coverage comparison results. LitCovid tracks four times more articles on Long COVID (9,084 v.s. 2,261) and ~90% of Long COVID articles in Clinical Queries are already in LitCovid. The articles that are unique in Clinical Queries are either preprints – which are not tracked by LitCovid (3) – or false positives. That is, they contain Long COVID-related keywords but do not discuss Long COVID, e.g., PMID 35455319 mentions "Long-Term COVID-19" in the context of "Short-Term and Long-Term COVID-19 Pandemic Forecasting."

*COVID-19 variant and vaccine annotations.* Both named entity recognition and normalization modules achieved robust performance with F1-scores over 0.98. We will continue to monitor and update the method for new variants and vaccines. They are now used as filters in the interface to improve the search experience, which we will describe below.

*Document triage and topic annotations*. The document classification method achieved an F1-score of 0.96 for classifying whether an article is on COVID-19. In addition, the topic annotation method achieved an F1-score of 0.92. It had a better overall performance than the results reported by the LitCovid challenge overview from 80 system submissions worldwide (7) and only required ~20% of the inference time of the previous topic annotation model in the LitCovid production environment.

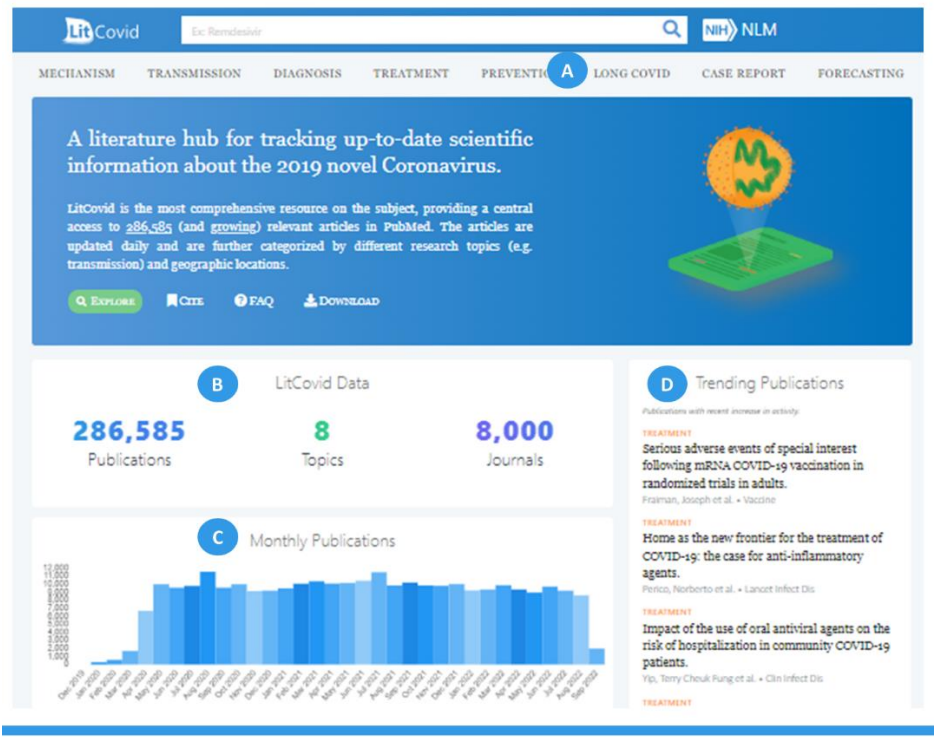

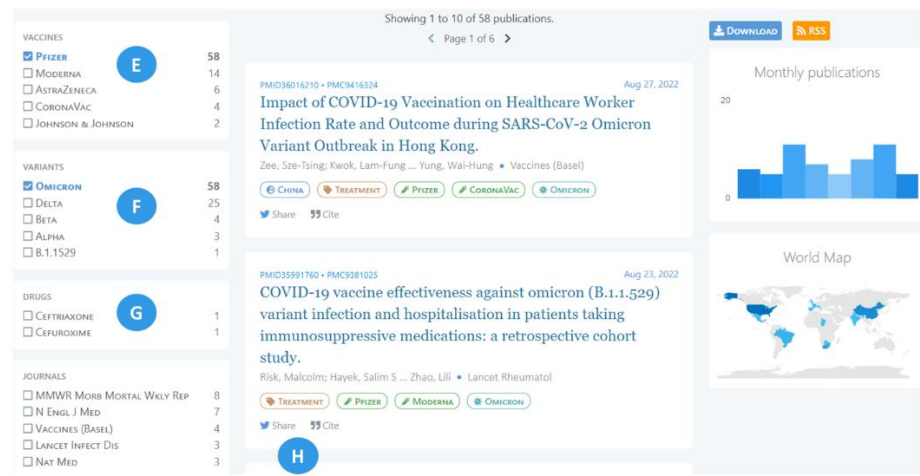

Figure 4. The updated LitCovid interface (screenshot in 10th September, 2022).

## Improvements in the user interface and other search functions

We further improved the LitCovid interface to maximize the user experience. Figure 4 demonstrates the latest interface (screenshot on 10th September 2020). First, we added the newly annotated "Long Covid" topic (A) to the toolbar at the top for more convenient access. Second, we added a brief overview widget (B) to the homepage to present the main statistics of our database content: number of publications, journals, and topics. Third, we updated the growth of publications histogram (C) by monthly to track a longer period of the COVID-19 pandemic. Finally, we further presented a more informative "Trending Publications" widget (D), allowing users to quickly grasp the current state of COVID19 research. The trending articles are generated by filtering publications trending in PubMed by their presence in the LitCovid database.

In addition, we enhanced the document summary page to improve the search experience. First, we added two new filters, vaccines (E) and variants (F), reflecting the latest change in tracking COVID-19 vaccines and variants as mentioned above. Second, we replaced the "chemical" filter by a more specific "drugs" filter (G). Users can search articles across multiple filters. For instance, the bottom part of the figure shows an example of the articles co-mentioning the Omicron variant and Pfizer vaccine. For each publication, we also added the functions to share and cite the publication (H).

## CONCLUSION

LitCovid has continued its central role in accumulating the latest knowledge on the COVID-19 literature. In this article, we describe the continuing efforts on improving LitCovid over the past two years in response to the latest status of the COVID-19 pandemic, such as tracking Long COVID-related articles and annotating the latest variants and vaccines. LitCovid has been updated daily for over two years and the entire data collection is freely available to the community. Possible research directions for LitCovid development include the integration of full-length articles in PubMed Central and the development of COVID-19-specific language models for downstream curation tasks and text mining applications. We also look forward to user feedback to guide further enhancements.

## DATA AVAILABILITY

LitCovid is free and open to all users and there is no login requirement. LitCovid can be accessed via https://www.ncbi.nlm.nih.gov/research/coronavirus/.

## FUNDING

Intramural Research Program of the National Library of Medicine, National Institutes of Health. Funding for open access charge: Intramural Research Program of the National Library of Medicine, National Institutes of Health.

*Conflict of interest statement*. None declared.